\documentclass{article}
\usepackage{spconf,amsmath,graphicx}
\usepackage{amssymb}
\usepackage[hidelinks]{hyperref}
\usepackage{tablefootnote}
\usepackage{multirow}
\usepackage{cite, array, booktabs}
\usepackage{subcaption}
\usepackage{epstopdf} 

\title{EEND-SS: Joint End-to-End Neural Speaker Diarization and Speech Separation for Flexible Number of Speakers}
\name{Soumi Maiti$^{1\star}$, Yushi Ueda$^{1\star}$, Shinji Watanabe$^1$, 
Chunlei Zhang$^2$, Meng Yu$^2$, Shi-Xiong Zhang$^2$, Yong Xu$^2$}
\address{
  $^1$Carnegie Mellon University, Pittsburgh, PA, USA, $^2$Tencent AI Lab, Bellevue, WA, USA
 }

\copyrightnotice{978-1-6654-7189-3/22/\$31.00~\copyright2023 IEEE}

\begin{document}
\ninept
\maketitle
\begin{abstract}
In this paper, we present a novel framework that jointly performs three tasks: speaker diarization, speech separation, and speaker counting. Our proposed framework integrates speaker diarization based on end-to-end neural diarization (EEND) models, speaker counting with encoder-decoder based attractors (EDA), and speech separation using Conv-TasNet. In addition, we propose a multiple 1×1 convolutional layer architecture for estimating the separation masks corresponding to a flexible number of speakers and a fusion technique for refining the separated speech signal with obtained speaker diarization information to improve the joint framework.
Experiments using the LibriMix dataset show that our proposed method outperforms the single-task baselines in both diarization and separation metrics for fixed and flexible numbers of speakers and improves speaker counting performance for flexible numbers of speakers. All materials will be open-sourced and reproducible in ESPnet toolkit\footnote{\url{https://github.com/espnet/espnet}}.

\end{abstract}

\renewcommand{\thefootnote}{\fnsymbol{footnote}}
\footnotetext[1]{The two authors contributed  equally to this paper.}

\section{Introduction}
Speech separation and speaker diarization are key technologies for various speech processing applications, including automatic speech recognition for multi-speaker speech mixtures such as meetings ~\cite{carletta2005ami, janin2003icsi} or parties~\cite{watanabe2020chime}. Speaker diarization is the task of estimating multiple speakers' speech activities (``who spoke when") from the input audio~\cite{park2022review}. On the other hand, speech separation is the task of separating each speaker from the input mixture audio. If we know the answer to ``who spoke when" beforehand, then it is reasonable to expect that we could separate the overlapped speech more efficiently and \textit{vice versa}. Thus, intuitively we can say that these two tasks: diarization and separation, are mutually related, and solving one problem would benefit the performance of the other. However, in most cases, it is not possible to obtain either of the information in advance. Additionally, if the number of speakers in the speech mixture is unknown, the two tasks become even more challenging.

Traditional clustering-based diarization systems~\cite{sell2014speaker,shum2013unsupervised} assume that only one speaker is active at a time. Thus they cannot handle speaker overlapped data and hence are less beneficial for speech separation.
Such clustering-based diarization systems are not end-to-end models as well.
Conversely, fully end-to-end neural diarization (EEND)~\cite{Fujita2019, Fujita2019asru, liu2021end} systems can handle speaker overlap by training with the speaker overlap data.
One drawback of EEND is the number of speakers has to be known and fixed beforehand. Several techniques have been proposed for EEND with a variable number of speakers, such as using the maximum number of speakers in the mixture~\cite{maiti2021end} or iteratively extracting one speaker activity at a time using a conditional speaker chain rule~\cite{fujita2020neural}. The most straightforward work is EEND with Encoder-Decoder-based Attractor calculation (EEND-EDA)~\cite{horiguchi2020Interspeech}. EEND-EDA counts speakers as a subtask within diarization using LSTM encoder-decoder based attractors.

On the other hand, several works in speech separation are proposed to handle a variable number of speakers. Some of the key approaches include: recursively separating the speakers one by one~\cite{takahashi2019recursive, kinoshita2018icassp, shi2020nips}; inferring the number of speakers before the separation, and then selecting the model corresponding to the number of speakers~\cite{Junzhe2021}. Another approach first separates using the model for the largest possible number of speakers and then uses speech detection on separated signals to select the model for the detected number of speakers~\cite{nachmani2020voice}.

Even though speaker diarization and speech separation are often used together as building blocks in speech systems, their optimal order is not fixed, and this order varies with the scenario and dataset~\cite{watanabe2020chime, chen2020css, Desh2021slt}. This different ordering issue suggests that we should solve these two tasks jointly. So, our solution is to unify these tasks in a single neural network and jointly train it with multi-task learning so that both tasks can benefit from each other. Some previous work shows that joint modeling with voice activity detection (VAD) improves speaker diarization~\cite{takashima2021end}, target speech separation~\cite{lin2021sparsely}, and speech enhancement tasks~\cite{tan2021speech}. Online Recurrent Selective Attention Network (RSAN)~\cite{Thilo2019, kinoshita2020tackling} proposes to jointly model speaker counting, diarization, and separation. RSAN focuses on one speaker's separation iteratively. By doing so, they inherently learn each speaker's activity information. Though the motivation is similar, our key contribution is that our proposed model optimizes speaker counting, diarization, and separation directly in a multitasking fashion. The proposed model does not require an iterative process, which could be affected by error propagation. 

More precisely, this paper proposes a novel framework: \textit{Joint End-to-End Neural Speaker Diarization and Separation (EEND-SS)}, which integrates end-to-end speaker diarization and speech separation tasks. The proposed framework is generalizable to use any speech separation or end-to-end speaker diarization technique. We chose Conv-TasNet~\cite{Yi2019tasnet} as a separation method as it is a very well-known separation model, and
EEND-EDA~\cite{horiguchi2020Interspeech} as a diarization model due to its end-to-end framework and overlap handling.
EEND-SS integrates both tasks into one network that minimizes speech separation, speaker diarization, and speaker counting errors directly and with multitask learning. Additionally, we propose two improvements to enhance integration between the tasks.
First, we propose multiple 1$\times$1 convolutional layers that can estimate separation masks corresponding to a input mixture with variable number of speakers, and we estimate the number of speakers from the diarizaion branch. Second, we propose a fusion technique for refining the separated speech signals with diarization branch learned speech activity. 
Experimental results show that EEND-SS can improve separation and diarization performances using 2-speaker and 3-speaker datasets for both fully and sparsely overlapped datasets. EEND-SS also improves speaker counting performance when used with a variable number of speakers, shown with a mix of 2 and 3 speakers. 

\section{Conventional methods}
\label{sec:method}
In this section, we introduce the conventional speaker diarization, speaker counting, and speech separation methods.
Let $\mathbf{x} \in \mathbb{R}^{1 \times T}$ be a single-channel $T$-length input speech mixture of $C$ speakers.
Then, input speech mixture $\mathbf{x}$ in an anechoic condition can be represented as\footnote{Ideally, if $s_c$ provides the complete 0 energy in the silence region, then we do not need $y_c$}:
\begin{equation}
\mathbf{x}=\sum_{c=1}^{C}\mathbf{y}_c \mathbf{s}_c+\mathbf{n}
\label{eq:x}
\end{equation}
Here, $\mathbf{s}_c \in \mathbb{R}^{1 \times T}$ is the source speech signal of speaker $c$.
${\mathbf{y}_c \in \{0,1\}^{T}}$ is the speech activity of speaker $c$, where $y_{c,t}=1$ indicates that speaker $c$ is speaking at time $t$ otherwise $y_{c,t}=0$.
$\mathbf{n}\in \mathbb{R}^{1 \times T}$ is a noise signal.

Speaker diarization estimates the speaker label sequence $\hat{Y}=  \{\hat{y}_{c,t}\} \in \{0,1\}^{C \times T}$, speech separation task predicts the separated speech signals $\mathbf{\hat{s}}_1,\cdots,\mathbf{\hat{s}}_C \in \mathbb{R}^{1 \times T}$, and speaker counting generates the number of speakers $\hat{C}$, given $\mathbf{x}$.

\subsection{End-to-end Speaker Diarization Module}\label{ssec:eend} (EEND)~\cite{Fujita2019,Fujita2019asru,liu2021end} estimates multiple speaker's activities simultaneously from input mixture using a single neural network with permutation invariant training (PIT) loss. EEND predicts speaker activity as binary multi-class labels $\hat{Y} \in \{0,1\}^{C \times T}$. In contrast to clustering-based diarization techniques, EEND can model overlapped speech by setting $\hat{y}_{c_1,t}=1$ and $\hat{y}_{c_2,t}=1$ if two speakers $c_1$ and $c_2$ are active at the same time $t$.  

Given log-mel filterbank ($\mathsf{LMF}$) features from input mixture, stack of transformer encoder ($\mathsf{TrfEnc}$) layers learns $D$-dimensional diarization embedding $\mathbf{e}_t \in \mathbb{R}^D$ as:
\begin{equation}\label{eq:diar_emb}
    \{\mathbf{e}_t\}^T_{t=1} = \mathsf{TrfEnc}(\mathsf{LMF}(\mathbf{x})) \in \mathbb{R} ^{D \times T}
\end{equation}
Diarization embeddings are then mapped to speaker activity probabilities $\{\mathbf{p}_{t}\}_{t=1}^{T}  \in (0,1)^{C \times T}$ with a linear layer and an element-wise sigmoid function $\sigma(\cdot)$, i.e.
\begin{equation}\label{eq:diar_prob}
    \mathbf{p}_{t} =\sigma (\mathbf{W}\mathbf{e}_t+ \mathbf{b})
\end{equation}
where $\mathbf{W} \in \mathbb{R}^{C \times D}$ and $\mathbf{b} \in \mathbb{R}^C$.
EEND is trained with the permutation invariant training loss between the speaker activity probabilities and the ground-truth speaker activity labels. Training loss for diarization ($\mathcal{L}_{\text{diar}}$) is defined as:
\begin{equation} \label{eq:loss_diar}
\mathcal{L}_{\text{diar}}=\underset{\phi \in \Phi(C)}{\min} \underset{t}{\sum} \mathsf{BCE}(\mathbf{y}_{t}^{\phi},\mathbf{p}_{t})
\end{equation}
Here, $\Phi(C)$ is a set of all possible permutations of $(1,\cdots,C)$, and $\mathbf{y}_t^{\phi}$ is a vector representation of permuted reference speaker labels. $\mathsf{BCE}(\cdot, \cdot)$ is the binary cross entropy loss.
Finally the speaker activity label ($\hat{y}_{c,t}$) is predicted for each speaker $c$ at time $t$, as introduced in the preliminary part of Section \ref{sec:method}, by applying a threshold $p_{c,t}$. 

\subsection{Speaker Counting Module based on EEND-EDA}\label{ssec:edad}
One drawback of EEND is that the number of speakers $C$ has to be fixed in advance.
To mitigate this difficulty, EEND with Encoder-Decoder Attractor (EEND-EDA)~\cite{horiguchi2020Interspeech} was proposed, which handles a flexible number of speakers by predicting speaker existence with attractor existence probability. With the assumption of the maximum possible number of speakers $C$, attractor existence labels are defined as $\mathbf{l}:=[l_1, \cdots l_{(C+1)} \in \{0,1\}]$
\begin{equation}\label{eq:attractor_label}
    l_c = \begin{cases}
    1 &(c \in \{1 \cdots C\}\\
    0 &(c= C+1)
    \end{cases}
\end{equation}

EDA takes $T$-length diarization embedding sequences $\{\mathbf{e}_t\}_{t=1}^{T}$, as introduced in Eq. \eqref{eq:diar_emb}, as input and calculates flexible number of attractor vectors $A := [\mathbf{a}_1, \cdots \mathbf{a}_C] \in \mathbb{R}^{D \times C}$. 
Then speaker activity probability $\mathbf{p}_{t}$, previously discussed in Eq. \eqref{eq:diar_prob}, is then reformulated as:
\begin{equation}
    \mathbf{p}_{t} =\sigma ({A}^{\top}\mathbf{e}_t).
    \label{eq:pt_eda}
\end{equation}
where ${\top}$ denotes transpose operation. Attractor existence probabilities $q_c \in (0,1)$ are calculated with a linear layer and a sigmoid function applied to $\mathbf{a}_c$. During training, the oracle number of speaker $C$ is known, inference number of speakers $\hat{C}$, as introduced in the preliminary part of Section \ref{sec:method}, is estimated by using $q_c$. 

The training objective of the attractor existence probabilities ($\mathcal{L}_{\text{exist}}$) is defined as:
\begin{equation}
    \mathcal{L}_{\text{exist}}=\frac{1}{C+1}\mathsf{BCE}(\mathbf{l},\mathbf{q}),
    \label{eq:loss_exist}
\end{equation}
 and  $\mathbf{q}:=[q_1,\cdots,q_{C+1}]$. During inference, $\hat{C}$ is estimated by counting the first $\hat{C}$ attractor existence probabilities $q_c$ that are larger than a given threshold.

\subsection{Speaker Separation Module}\label{ssec:sep}
As discussed in the preliminary part of Section \ref{sec:method}, the speech separation task estimates source $\{\mathbf{\hat{s}}_c\}_{c=1}^C$ from the input mixture $\mathbf{x}$.
We use Convolutional Time-domain Audio Separation Network (Conv-TasNet)~\cite{Yi2019tasnet} as the speech separation method in this paper, though other separation methods could be used as well.

\begin{figure}[t]
  \centering
  \includegraphics[width=0.98\linewidth, trim={0 5 9 2},clip]{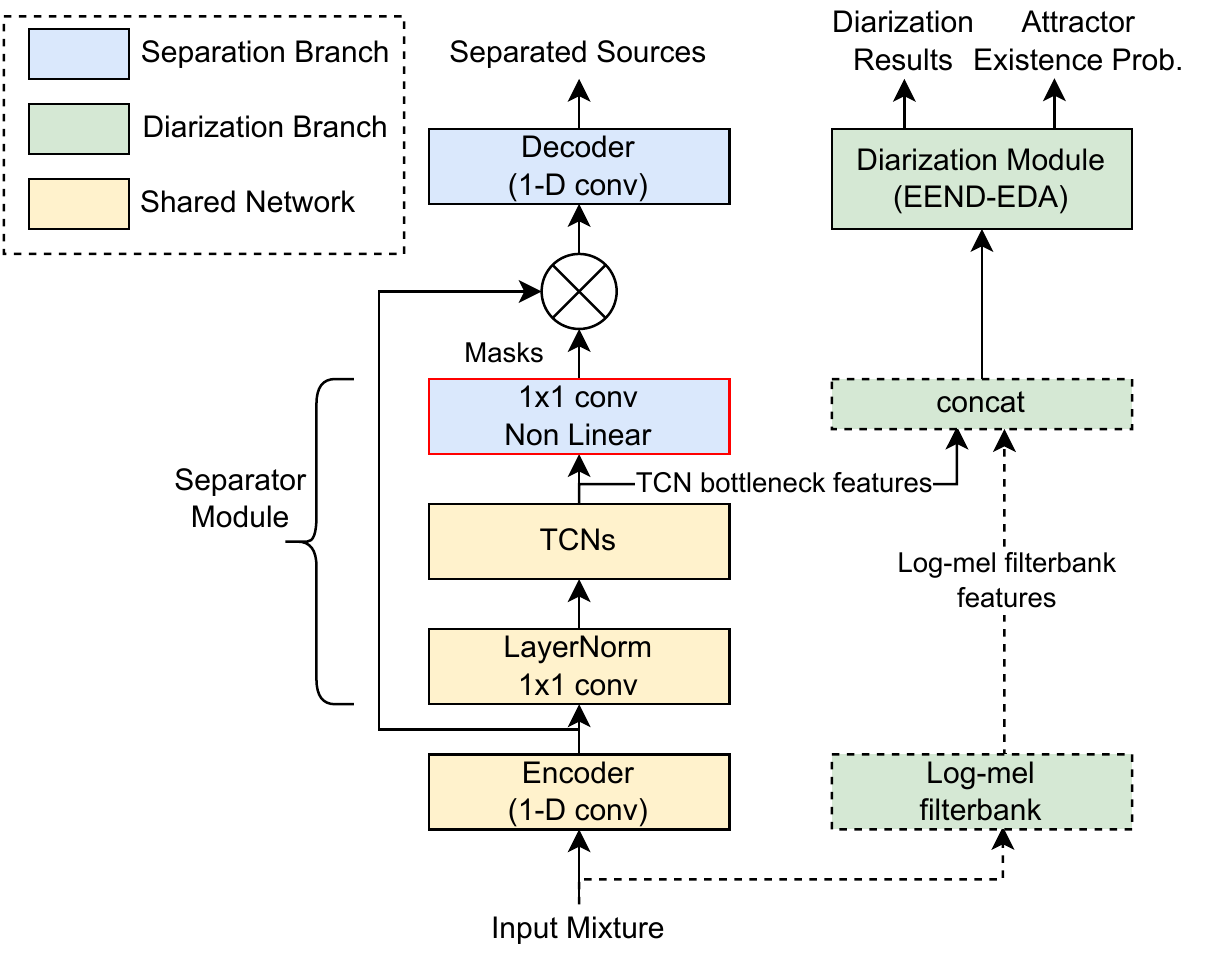}
  \caption{Overall structure of the proposed model (EEND-SS).}
  \label{fig:model}
  \centering
  \includegraphics[width=0.9\linewidth, trim={0 5 0 2},clip]{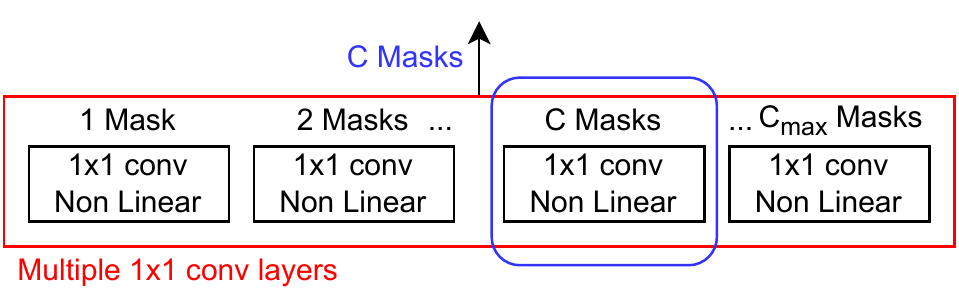}
  \caption{Multiple 1$\times$1 convolutional layer architecture. }
  \label{fig:conv1x1}
\end{figure}

Conv-TasNet~\cite{Yi2019tasnet} is one of the most well-known speech separation methods that separate the audio signal in the time domain. Conv-TasNet consists of three fully convolutional modules: encoder, decoder, and separator.
%
It uses a convolution encoder $\mathsf{ConvEnc}(\cdot)$ to encode the input audio signal $\mathbf{x}$ to $N$-dimensional representations $H = \{\mathbf{h}_{t}\}_{t=1} ^{T} \in \mathbb{R}^{N \times T}$ 
\begin{equation}\label{eq:sep_feat}
    \{\mathbf{h}_{t}\}_{t=1} ^{T} = \mathsf{ConvEnc}(\mathbf{x})
\end{equation}
$\mathsf{ConvEnc}(\cdot)$ consists of a 1-D convolution layer followed by a ReLU.
In the separator, $H$ is processed by a global layer normalization and a 1$\times$1 convolutional layer followed by repeated temporal convolutional network (TCN) blocks.
Each TCN block is composed of stacked 1-D dilated convolutional layers with exponentially increasing dilation factors, and such blocks are repeated.  We refer to the repeated stacked blocks of TCN modules simply as $\mathsf{TCNs}$ in the Fig.~\ref{fig:model}. 
$\mathsf{TCNs}$ outputs $B$-dimensional embeddings $\mathbf{e}_{t}^{\text{tcn}} \in \mathbb{R}^{B}$:
\begin{equation}
    \{\mathbf{e}_{t}^{\text{tcn}}\}_{t=1} ^{T} = \mathsf{TCNs}( \mathsf{1x1Conv}(\mathsf{LayerNorm}((H))
\end{equation}
$\mathbf{e}_{t}^{\text{tcn}}$ is also referred as \textit{TCN bottleneck features} and $B$ is bottleneck dimension.
The separator module estimates masks $\mathbf{m}_{c,t} \in [0,1]^{N} $:
\begin{equation}\label{eq:mask}
    \{\mathbf{m}_{c,t}\}_{c=1}^{C} =  \sigma (\mathsf{1x1Conv}(\mathsf{PReLU}((\mathbf{e}_{t}^{\text{TCN}})))
\end{equation}
The representation for each source $\mathbf{d}_{c,t} \in \mathbb{R}^{ N}$ is computed as:
\begin{equation}
    \mathbf{d}_{c,t} = \mathbf{h}_{t} \odot \mathbf{m}_{c, t} 
\end{equation}
where $\odot$ denotes the element-wise multiplication.

Separated audio signals $\mathbf{\hat{s}}_c$, as introduced in the preliminary part of Section \ref{sec:method}, are estimated with a $\mathsf{Decoder}$ as: 
\begin{equation}
    \mathbf{\hat{s}}_c =\mathsf{Decoder} (\mathbf{d}_{c})
    \label{eq:s_decoder}
\end{equation}
$\mathsf{Decoder}$ consists of a 1-D transposed convolutional layer.
Conv-TasNet is trained with the SI-SDR~\cite{leroux2019icassp} loss defined as:
\begin{equation}
    \mathcal{L}_{\text{SI-SDR}}=-10\text{log}_{10}\frac{
    \left\| \frac{\left\langle \mathbf{\hat{s}},\mathbf{s} \right\rangle\mathbf{s}}{\left\| \mathbf{s} \right\|^2} \right\|^2
    }{\left\|\mathbf{\hat{s}} - \frac{\left\langle \mathbf{\hat{s}},\mathbf{s} \right\rangle\mathbf{s}}{\left\| \mathbf{s} \right\|^2} \right\|^2
    }.
    \label{eq:loss_si_sdr}
\end{equation}

\section{Proposed Joint Speaker Diarization and Speech Separation (EEND-SS)}\label{subsec:proposed}

\subsection{Overall Structure}
Our proposed model, \textit{Joint End-to-End Neural Speaker Diarization and Separation (EEND-SS)}, performs three tasks: speaker diarization, speech separation, and speaker counting. EEND-SS takes speech mixture $\mathbf{x}$ as input. The proposed model consists of a shared network between all three tasks, followed by two separate branches: one for speech separation and another for diarization. Speaker counting is integrated as a subtask of the diarization module, similar to the original proposed EEND-EDA. 

Fig.~\ref{fig:model} shows the overall structure of EEND-SS. In Fig.~\ref{fig:model} shared branch of EEND-SS are indicated as yellow blocks, the diarization-only branch as green, and the separation-only branch as blue. Input mixture up to \textit{TCN bottleneck features} as defined in the previous section~\ref{ssec:sep} are shared between both networks. Diarization branch uses learned TCN features $\mathbf{e}_{t}^{\text{tcn}}$ as input acoustic features. i.e. Eq.~\eqref{eq:diar_emb} is reformulated as:
\begin{equation}\label{eq:diar_emb_joint}
    \{\mathbf{e}_t \}_{t=1} ^T = \mathsf{TrfEnc}(\{\mathbf{e}_{t}^{\text{tcn}} \}_{t=1} ^T)
\end{equation}
Optionally, LMF features concatenated with TCN bottleneck features can be passed to the EEND-SS diarization module (shown with dotted lines in Fig.~\ref{fig:model}) as:
\begin{equation}\label{eq:diar_emb_joint_concat}
    \{\mathbf{e}_t  \}_{t=1} ^T= \mathsf{TrfEnc}(\mathsf{Concat}(\{\mathbf{e}_{t}^{\text{tcn}} \}_{t=1} ^T, \mathsf{LMF}(\mathbf{x}))
\end{equation}
EEND-EDA uses a subsampling layer on the LMF features, and similar subsampling is also applied on $\mathbf{e}_{t}^{\text{tcn}}$. In the optional case, LMF is concatenated with $\mathbf{e}^{\text{tcn}}$ after the subsampling.

The separation branch follows the same architecture as defined in the previous section~\ref{ssec:sep}. We extend the mask estimation layer in the separator for a flexible number of speaker handling, which is described in detail below.

\subsection{Multiple 1$\times$1 Convolutional Layers}\label{ssec:1x1conv}
In EEND-SS, we use multiple 1$\times$1 convolutional layers to extend the mask generation capability to a different number of speakers. In Conv-TasNet, last 1$\times$1 convolutional layer generates $C$ masks $\{\mathbf{m}_{c,t}\}_{c=1}^{C}$ corresponding to the fixed predetermined number of speakers $C$, as defined in Eq.~\eqref{eq:mask}.

Instead, in EEND-SS we use $C_{\text{max}}$ number of 1$\times$1 convolutional layers, where $C_{\text{max}}$ is the maximum possible number of speakers. Each 1$\times$1 convolutional layer estimates set of masks corresponding to a different number of speakers, starting from $1$ to $C_{\text{max}}$. For example $k$-th 1$\times$1 convolutional layer ($\mathsf{1x1 Conv}_k$) learns a set of $k$ masks, and Eq.~\eqref{eq:mask} is reformulated as:
\begin{equation}\label{eq:multiple1x1}
    \{\mathbf{m}_{c,t}\}_{c=1}^{k}= \mathsf{1x1Conv}_k (\mathbf{e}_{t}^{\text{TCN}}),
\end{equation}
where $k=1,\cdots ,C_{\text{max}}$. We skipped $\mathsf{PReLU}$ and $\sigma$ here from Eq.~\eqref{eq:multiple1x1} for simplification.
Multiple 1$\times$1 convolutional layer architecture is also shown in Fig.~\ref{fig:conv1x1}. Though $C_\text{max}$ 1$\times$1 convolutional layers are used, at one time, only one layer is chosen based on the number of speakers. During training, the oracle number of $C$ speakers is used,i.e., $k=C$. During inference, we use $k=\hat{C}$, the speaker number estimated by the diarization branch.

This architecture is similar to multi-decoder DPRNN~\cite{Junzhe2021} in terms of selecting the network corresponding to the estimated number of speakers. However, while multi-decoder DPRNN switches the whole decoder, EEND-SS only switches a single layer and shares the decoder structure. 
Thus, the decoder in EEND-SS is trained using the input mixture with various numbers of speakers. 
This architecture is thought to be efficient, especially when the training samples including a specific number of speakers are scarce. 
In multiple 1$\times$1 convolutional layer architecture, the maximum number of speakers that the model can handle is bound to the number of multiple 1$\times$1 convolutional layers $C_{\text{max}}$. 
However, in practice, we can handle an arbitrary number of speakers by setting $C_{\text{max}}$ to a sufficiently large number.
Note that since the unused 1$\times$1 layers will not interfere with the rest of the network, we can safely set $C_{\text{max}}$ to a large number without hurting the performance.

\subsection{Fusion of Speech Activity and Separated Signals} 
In EEND-SS, during training, the two separate branches have only the shared network as mutual interaction. However, we use information predicted from the diarization branch during inference for refining separation. One example of such information is the predicted number of speakers, which is used for selecting the corresponding 1$\times$1 convolutional layer as mentioned in the previous section~\ref{ssec:1x1conv}. Another information we use is predicted speaker activity from the diarization branch. We use speech activity probabilities $\mathbf{p}$ estimated from the diarization branch and multiply with the separated speech signals $\mathbf{\hat{s}}$ from the decoder module in Eq.~\eqref{eq:s_decoder}.

For this step, we also need to find the corresponding speaker alignment between the separated speech signals and the diarization results since the output ordering of the speakers may differ for the two branches.  We find corresponding speaker alignment by selecting the combination that has maximum the sum of correlations between the amplitude of the separated speech signals and the posterior probabilities. Let $\mathbf{\hat{s}}^{\prime}$ be the separated speech signals after the fusion step, the fusion step can be formulated as follows:
\begin{align}
\label{eq:pp}
    \mathbf{\hat{s}}^{\prime} & =  \mathbf{\hat{s}} \odot \mathbf{p}^{\phi_{\text{max}}}, \\
    \mathbf{p}^{\phi_{\text{max}}} &:=  \underset{(\phi_1,\cdots ,\phi_C)\in \Phi(C)}{\text{argmax}}\sum_{c=1}^{C}r(\text{abs}(\mathbf{\hat{s}}),\mathbf{p}^{\phi}).
\end{align}
$r(\cdot,\cdot)$ denotes the correlation function, $\Phi(C)$ is as introduced in Eq. \eqref{eq:loss_diar}, and $\mathbf{p}^{\phi} := [p_{\phi_c,t} \in (0,1)|c=1,\cdots,C]$ is the permuted posterior probabilities.

The motivation behind this step is to reduce the background noise while the speaker is not present. There are some previous work~\cite{lin2021sparsely, takashima2021end, ochiai2020icassp} that shows such improvement using Voice Activity Detection with the separated speech signal.


\subsection{Training} 
The network is trained with a multi-task cost function
\begin{equation}
    \mathcal{L}=\lambda_{1}\mathcal{L}_{\text{SI-SDR}}+\lambda_{2}\mathcal{L}_{\text{diar}}+\lambda_{3}\mathcal{L}_{\text{exist}},
    \label{eq:loss}
\end{equation}
which is a weighted sum of $\mathcal{L}_{\text{SI-SDR}}$ in Eq.~\eqref{eq:loss_si_sdr}, $\mathcal{L}_{\text{diar}}$ in Eq.~\eqref{eq:loss_diar} and $\mathcal{L}_{\text{exist}}$ in Eq.~\eqref{eq:loss_exist}. $\lambda_1, \lambda_2, \lambda_3  \in \mathbb{R}_+$ are the weighting parameters that are chosen empirically.

\subsection{Inference} 
To handle a variable number of speakers, we utilize the following 2-pass inference procedure: 
(1) Obtain diarization probabilities $\mathbf{p}_t$ and the number of speakers $\hat{C}$ from the input speech mixture. 
(2) Select 1$\times$1 convolutional layer corresponding to $\hat{C}$ masks, then obtain separated speech signals
$\mathbf{\hat{s}}_{1},\cdots,\mathbf{\hat{s}}_{\hat{C}}$.
Optionally use $\mathbf{p}_t$ to further refine $\mathbf{\hat{s}}$ to $\mathbf{\hat{s}}^{\prime}$.

\section{Experiments}\label{sec:experiment}
\subsection{Experimental settings}
\subsubsection{Dataset} 
Since our target task is to solve diarization and separation tasks simultaneously, we need both ground truths, separated sources, and diarization labels to evaluate the performance objectively.
Though for diarization-only tasks, there are several real-world multiparty datasets available~\cite{carletta2005ami,horiguchi2021hitachi}, often separated sources are missing. 
Hence we use simulated conversation-like datasets to evaluate our model.
For the training and evaluation, we used the LibriMix\footnote{We used the groundtruth diarization labels available at \url{https://github.com/s3prl/LibriMix}}~\cite{cosentino2020librimix} and SparseLibriMix~\cite{cosentino2020librimix} datasets. LibriMix uses speech samples from LibriSpeech~\cite{panayotov2015librispeech} train-clean100/dev-clean/test-clean and the noise samples from WHAM!~\cite{Wichern2019WHAM} to generate mixtures for training/validation/testing. The dataset includes 58h/11h/11h of training/validation/testing sets for a two-speaker mixture (\textit{Libri2Mix}) and 40h/11h/11h for a three-speaker mixture (\textit{Libri3Mix}). We used an 8kHz sampling rate and the \textit{min} mode. For the SparseLibriMix testset, we use the original testset with six varying overlap conditions. For training with SparseLibriMix, we generated training data using $5k$ sentences per overlap condition and 90\%-10\% split for train-dev using released scripts with SparseLibrimix\footnote{https://github.com/popcornell/SparseLibriMix}.

\subsubsection{Configurations}
The model parameters used for the experiments are as follows: for the encoder and decoder, we set the kernel size to 16 and stride to 8. The number of 1-D convolutional layers in each TCN block is set to 8, and the TCN blocks are repeated 3 times. We also set $N=512$, $B=128$, and $D=256$, as introduced in previous section~\ref{ssec:sep}.
For the EEND-EDA, we use a 2-D convolutional layer with 1/8 sub-sampling as an input layer and 4-stacked Transformer encoders with 4 attention heads without positional encodings following EEND-EDA original setup.
We use 80-dimensional LMF converted from power spectra calculated with a frame length of 512 samples and a frameshift of 64 samples. 
We set the thresholds $\theta$ and $\tau$ for obtaining the diarization results and speaker counting to $0.5$. 
We empirically set the values of $\lambda_1$, $\lambda_2$ and $\lambda_3$ in Eq. \eqref{eq:loss} as $1.0$, $0.2$, $0.2$ respectively, unless otherwise noted. 
We use the same model parameters for Conv-TasNet, EEND-EDA and EEND-SS. 
We employed the Adam optimizer for training with a learning rate of $10^{-3}$ and a mini-batch size of 16. The learning rate was halved, and training was stopped if there was no improvement for 3 and 5 consecutive epochs, respectively.

\subsubsection{Evaluation Metrics} 
We report separation performance with three objective metrics: source-to-distortion ratio improvement ($\text{SDR}_i$(dB))~\cite{tr05}, scale-invariant source-to-distortion ratio improvement ($\text{SI-SDR}_i$(dB))~\cite{leroux2019icassp}, and short-time objective intelligibility (STOI)~\cite{Taal2010ICASSP}, and diarization performance with the diarization error rate (DER(\%))~\cite{DER}. When calculating the DER, collar tolerance of 0.0 sec and median filtering of 11 frames were used. We also report the Speaker Counting Accuracy (SCA(\%)) for speaker counting performance of the attractor module. 
\subsection{Results}
\subsubsection{Fixed Number of Speakers}
\begin{table}[t]
  \caption{Experimental results on Libri2Mix.  ``LMF" stands for Log-Mel Filterbank.}
  \label{tab:libri2mix}
  \centering
  \resizebox{0.47\textwidth}{!}{
  \begin{tabular}{ l| c c c c }
    \toprule
    \multicolumn{1}{c|}{\textbf{Method}} & 
    \multicolumn{1}{c}{\textbf{STOI}} ($\uparrow$) & 
    \multicolumn{1}{c}{$\mathbf{\textbf{SI-SDR}_{i}}$}($\uparrow$) &
    \multicolumn{1}{c}{$\mathbf{\textbf{SDR}_{i}}$}($\uparrow$) &
    \multicolumn{1}{c}{\textbf{DER}}($\downarrow$) \\
    \midrule
    Conv-TasNet  & 0.830 & 10.82 & 11.40 & --  \\
    EEND-EDA & -- & -- &  -- &  5.93  \\
    EEND-SS ($\lambda_{1}=0$) & -- & -- & -- & 5.26  \\
    \midrule
    EEND-SS                 & 0.838 & \textbf{11.20} & 10.57 & \multirow{2}{*}{\textbf{5.12}} \\
    \hspace{1em}+ Fusion    & 0.838 &  \textbf{11.20} & 10.67 & \\
    \hspace{1em}+ LMF       & 0.838 & 11.13 &  \textbf{11.71} & \multirow{2}{*}{5.02} \\
    \hspace{1em}+ LMF + Fusion & 0.838 & 11.13 & \textbf{11.71} & \\
    \bottomrule
  \end{tabular}
    }
\end{table}

\begin{table}[t]
  \caption{Experimental results on Libri3Mix.}
  \label{tab:libri3mix}
  \centering
  \resizebox{0.47\textwidth}{!}{
  \begin{tabular}{ l| c c c c }
    \toprule
    \multicolumn{1}{c|}{\textbf{Method}} & 
    \multicolumn{1}{c}{\textbf{STOI}}($\uparrow$) & 
    \multicolumn{1}{c}{$\mathbf{\textbf{SI-SDR}_{i}}$}($\uparrow$) &
    \multicolumn{1}{c}{$\mathbf{\textbf{SDR}_{i}}$}($\uparrow$) &
    \multicolumn{1}{c}{\textbf{DER}}($\downarrow$) \\
    \midrule
    Conv-TasNet &  0.721 & 7.94 & 8.73 &  --  \\
    EEND-EDA & -- & -- & --  & 8.81 \\
    EEND-SS ($\lambda_1=0$) & -- & -- & --  & 6.50 \\
    \midrule
    EEND-SS & 0.722 & 7.66 & 8.60 & \multirow{2}{*}{6.26} \\
    \hspace{1em}+ Fusion & 0.722 & 7.71 & 8.66 &  \\
    \hspace{1em}+ LMF & \textbf{0.723} &  8.39 & 8.96 & \multirow{2}{*}{\textbf{6.00}}\\
    \hspace{1em}+ LMF + Fusion & \textbf{0.723} &  \textbf{8.40} & \textbf{9.00} & \\
    \bottomrule
  \end{tabular}
  }
\end{table}
\begin{table}[t]
  \caption{Comparison of DERs on Libri2Mix max mode. ``SS Pretrained" indicates Self-Supervised Pretrained models.}
  \label{tab:superb}
  \centering
  \resizebox{0.4\textwidth}{!}{
  \begin{tabular}{ l| l| c }
    \toprule
    \multicolumn{1}{c|}{\textbf{Method}} &
    \multicolumn{1}{c|}{\textbf{Features}} &
    \multicolumn{1}{c}{\textbf{DER}}($\downarrow$) \\
    \midrule
    \multirow{4}{*}{EEND~\cite{yang21c_interspeech} }   & SS Pretrained & \\ 
     & \hspace{1em} wav2vec 2.0/HuBERT & \textbf{5.62}--6.08 \\
     & \hspace{1em} Others & 6.59--10.54\\
     & LMF & 10.05\\
    \midrule
    \multirow{2}{*}{EEND-SS} & TCN Bottleneck & 7.49 \\
     & TCN Bottleneck+LMF & \textbf{6.54} \\
    \bottomrule
  \end{tabular}
  }
\end{table}

First, we evaluated our method on fixed 2-speaker and 3-speaker conditions using Libri2Mix and Libri3Mix datasets, respectively. Both speaker diarization and speech separation performances are reported in Table~\ref{tab:libri2mix} and \ref{tab:libri3mix}. EEND-SS outperforms the baseline Conv-TasNet and EEND-EDA for both Libri2Mix and Libri3Mix datasets in all metrics. EEND-SS with multitasking loss also performs better than EEND-SS trained only on speaker diarization task (setting $\lambda_1=0$ in Eq.~\eqref{eq:loss}), for Libri2Mix and Libri3Mix datasets. 
Further performance gain for the speech separation is achieved by concatenating LMF described in Section \ref{subsec:proposed} and applying the fusion technique in Eq. \eqref{eq:pp}. 
Thus, we show the effectiveness of joint speech separation and speaker diarization based on the proposed method for fixed numbers of speakers.

\begin{figure}[!t]
    \centering
    \begin{subfigure}[b]{0.248\textwidth}
        \centering
        \includegraphics[width=1.\linewidth, trim={1mm 4mm 3mm 2mm},clip]{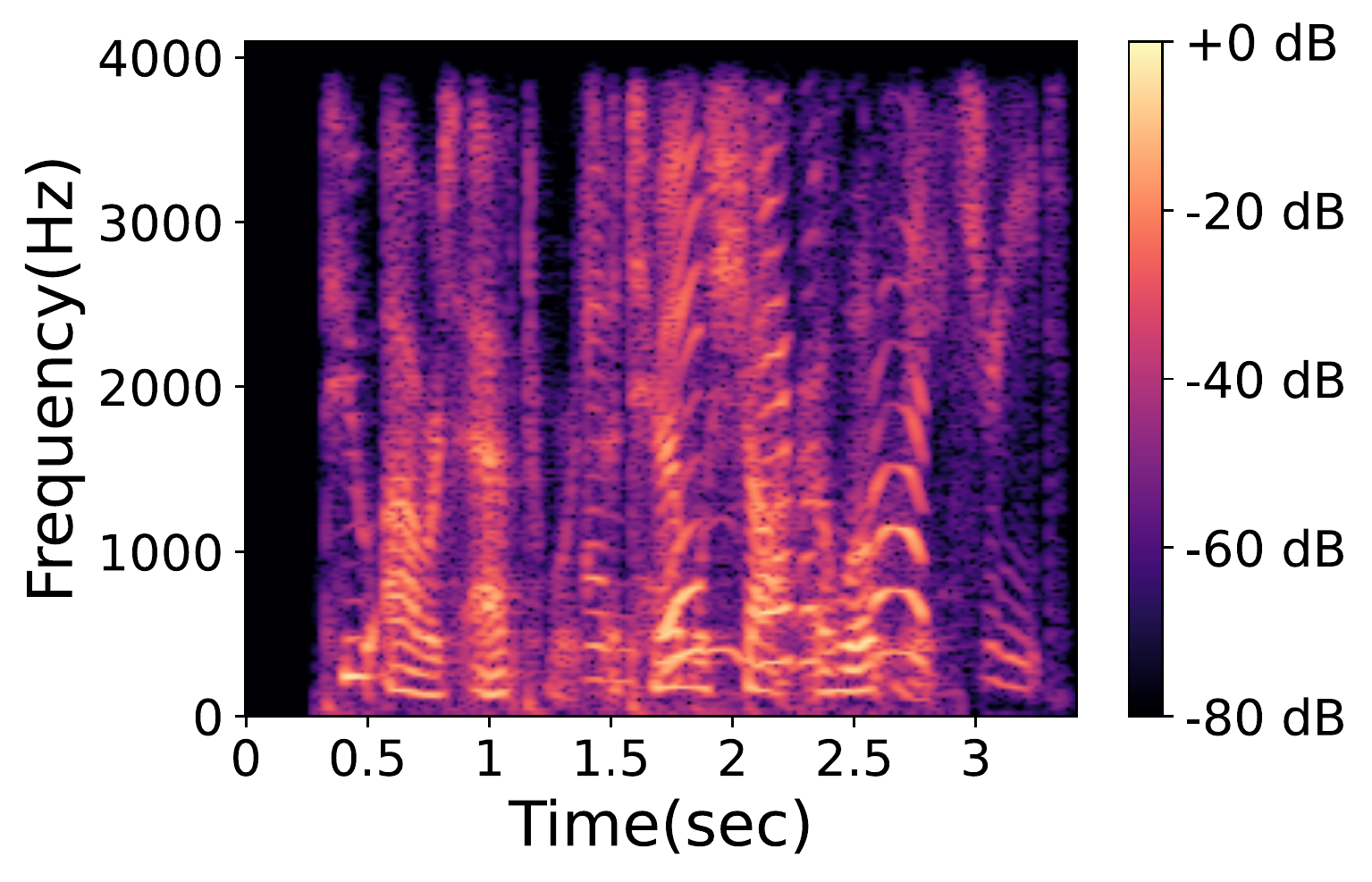}
        \caption{Input Mixture (Spk1+ Spk2)}
    \end{subfigure}%
    \newline
    \begin{subfigure}[b]{0.248\textwidth}
        \centering
        \includegraphics[width=1.\linewidth, trim={0 4mm 3mm 2mm},clip]{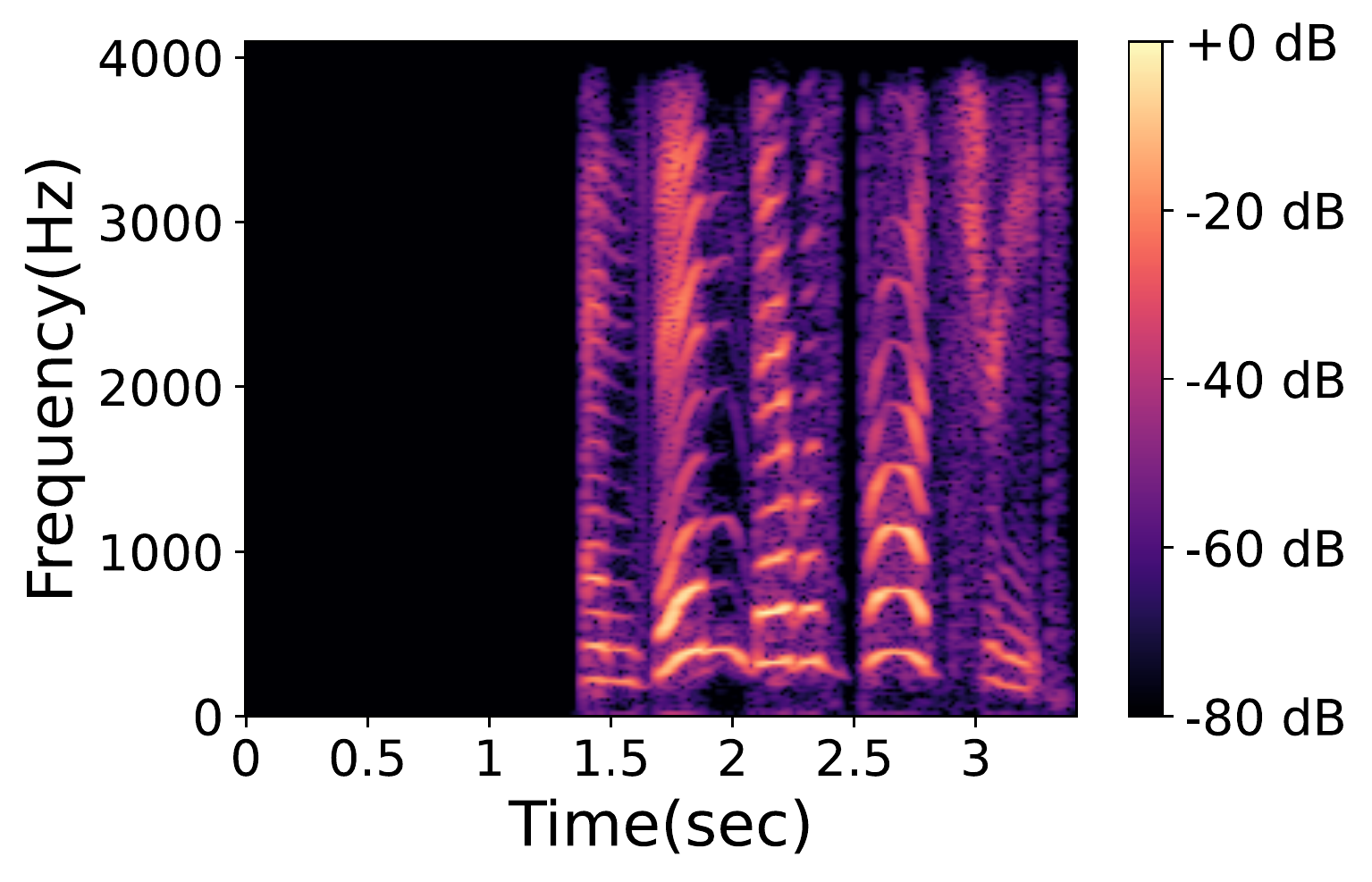}
        \caption{Ground Truth Spk1}
    \end{subfigure}%
    ~
    \begin{subfigure}[b]{0.248\textwidth}
        \centering
        \includegraphics[width=1.\linewidth, trim={0 4mm 3mm 2mm},clip]{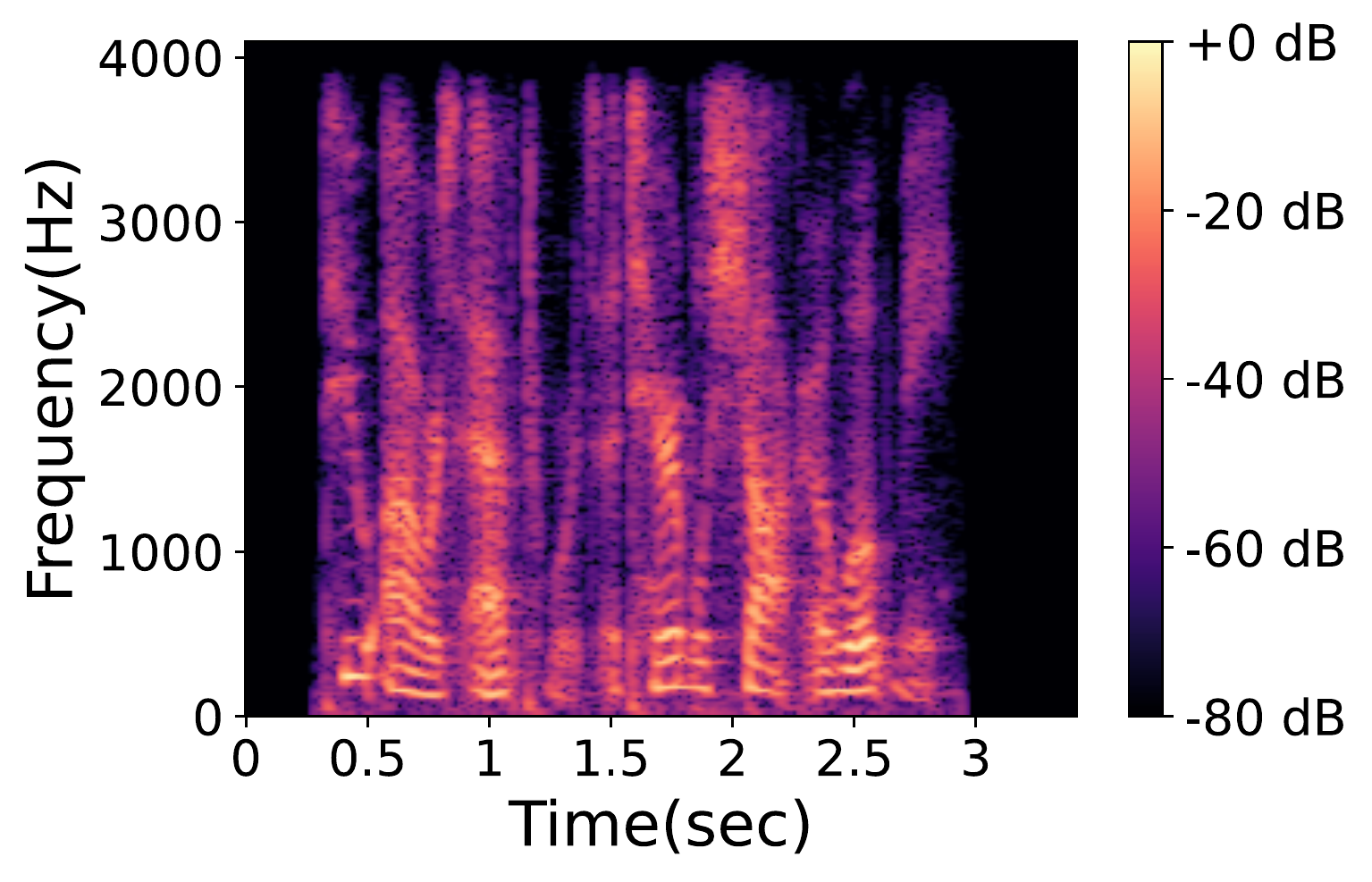}
        \caption{Ground Truth Spk2}
    \end{subfigure}%
    \newline
    \begin{subfigure}[b]{0.248\textwidth}
        \centering
        \includegraphics[width=1.\linewidth, trim={0 4mm 3mm 2mm},clip]{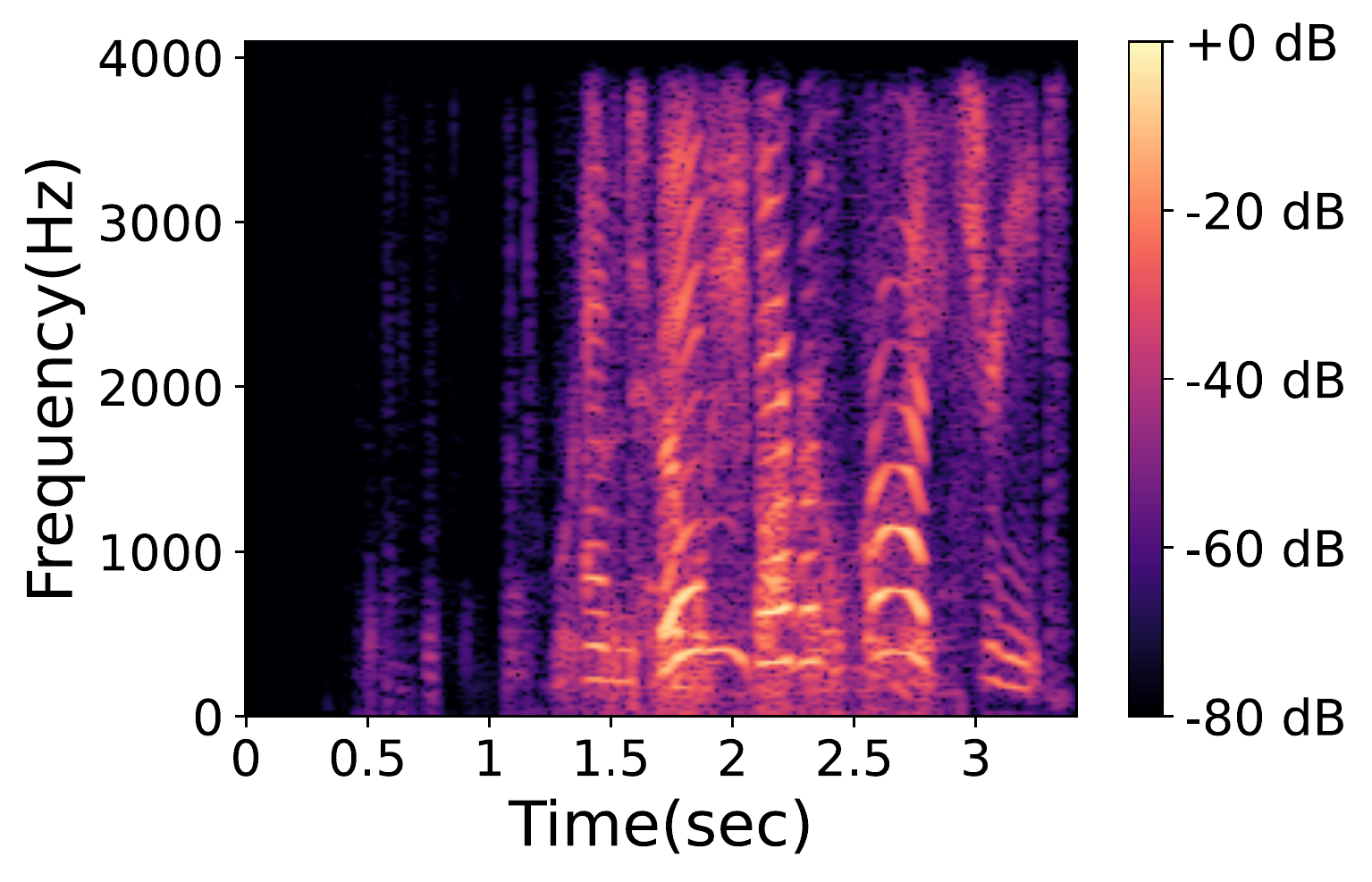}
        \caption{Recovered Spk1}
    \end{subfigure}%
    ~
    \begin{subfigure}[b]{0.248\textwidth}
        \centering
        \includegraphics[width=1.\linewidth, trim={0 4mm 3mm 2mm},clip]{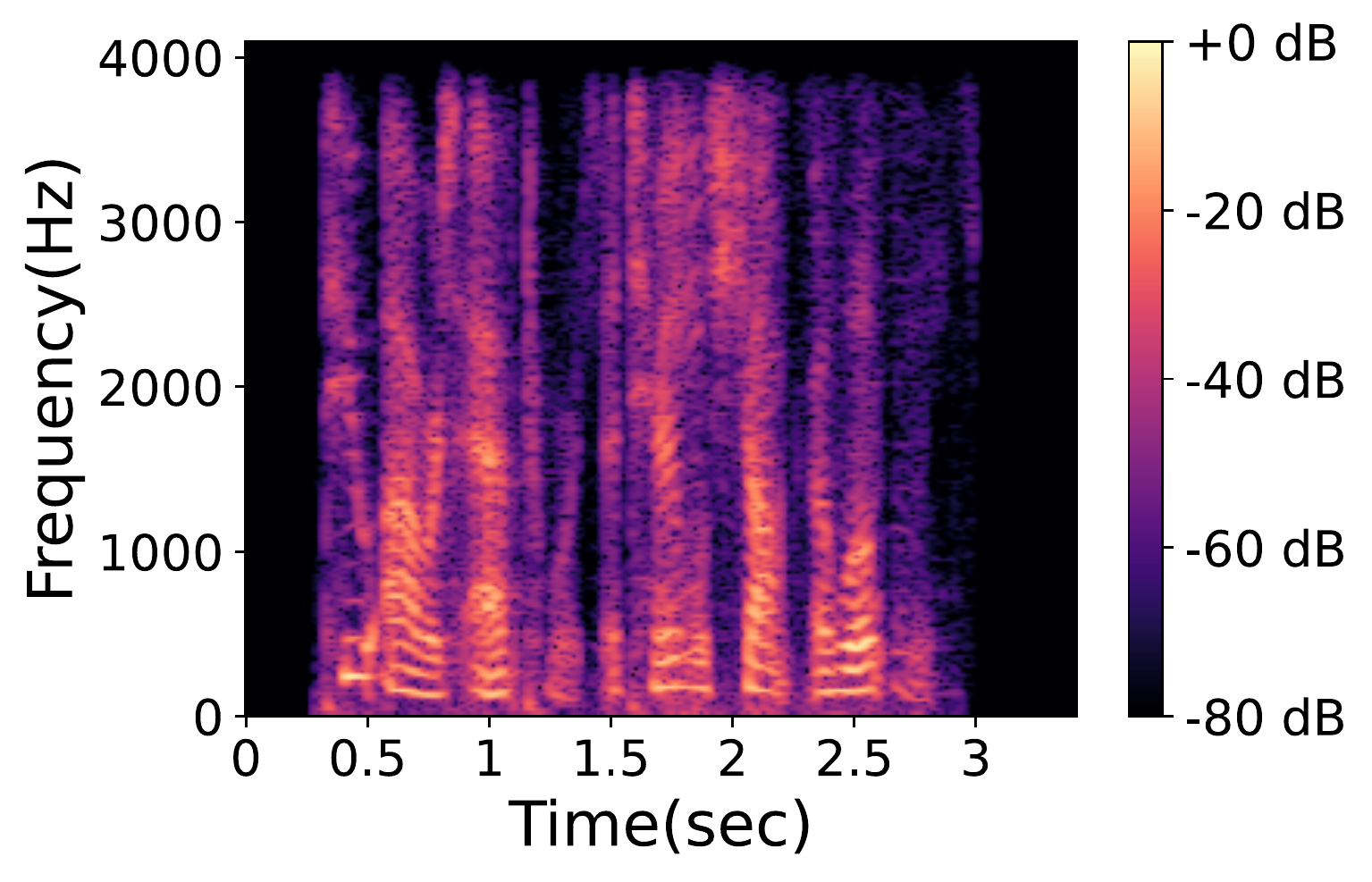}
        \caption{Recovered Spk2}
    \end{subfigure}%
    \newline
    \begin{subfigure}[b]{0.248\textwidth}
        \centering
        \includegraphics[width=1.\linewidth, trim={0 4mm 3mm 2mm},clip]{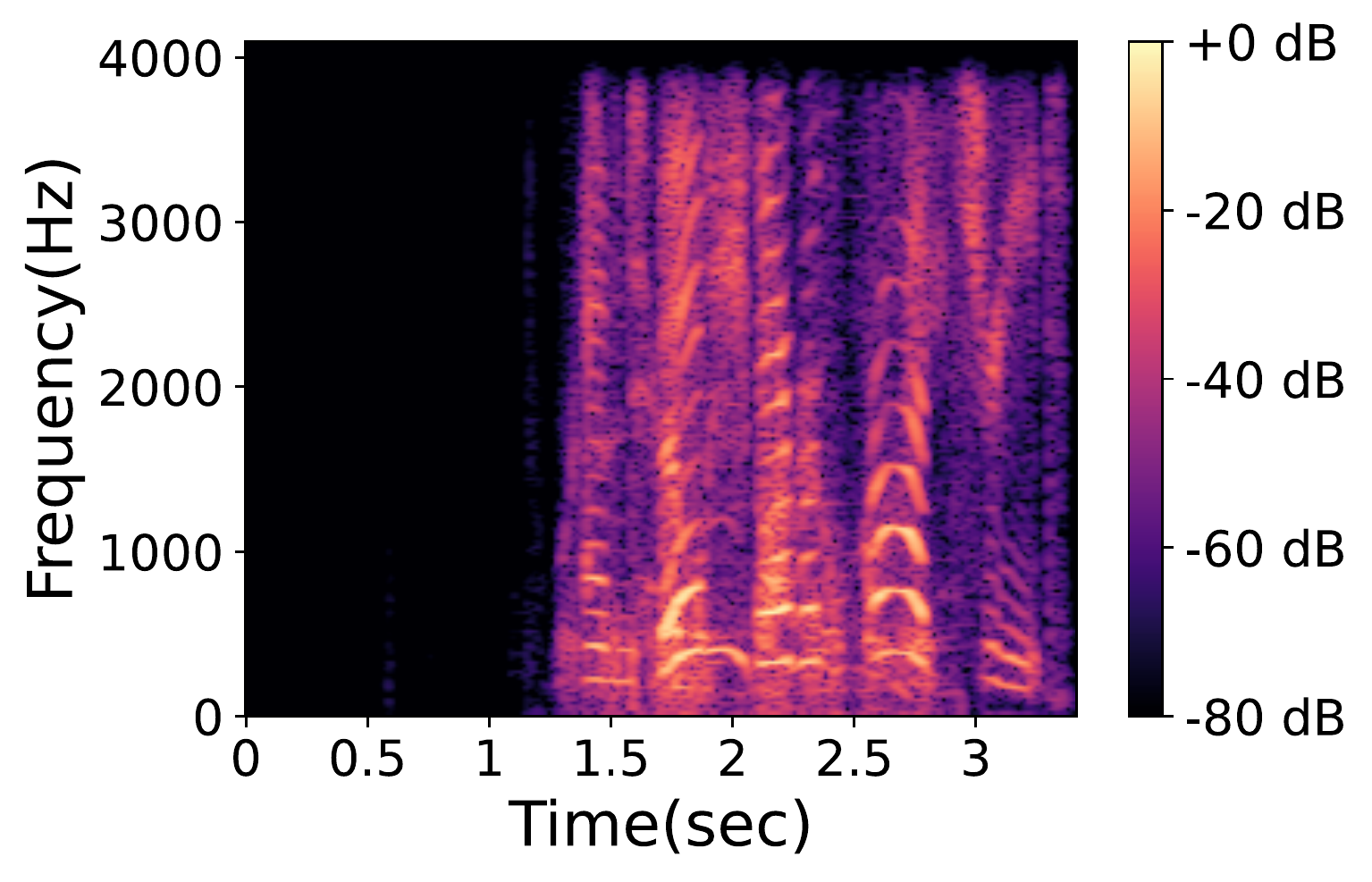}
        \caption{Recovered Spk1}
    \end{subfigure}%
    ~
    \begin{subfigure}[b]{0.248\textwidth}
        \centering
        \includegraphics[width=1.\linewidth, trim={0 4mm 3mm 2mm},clip]{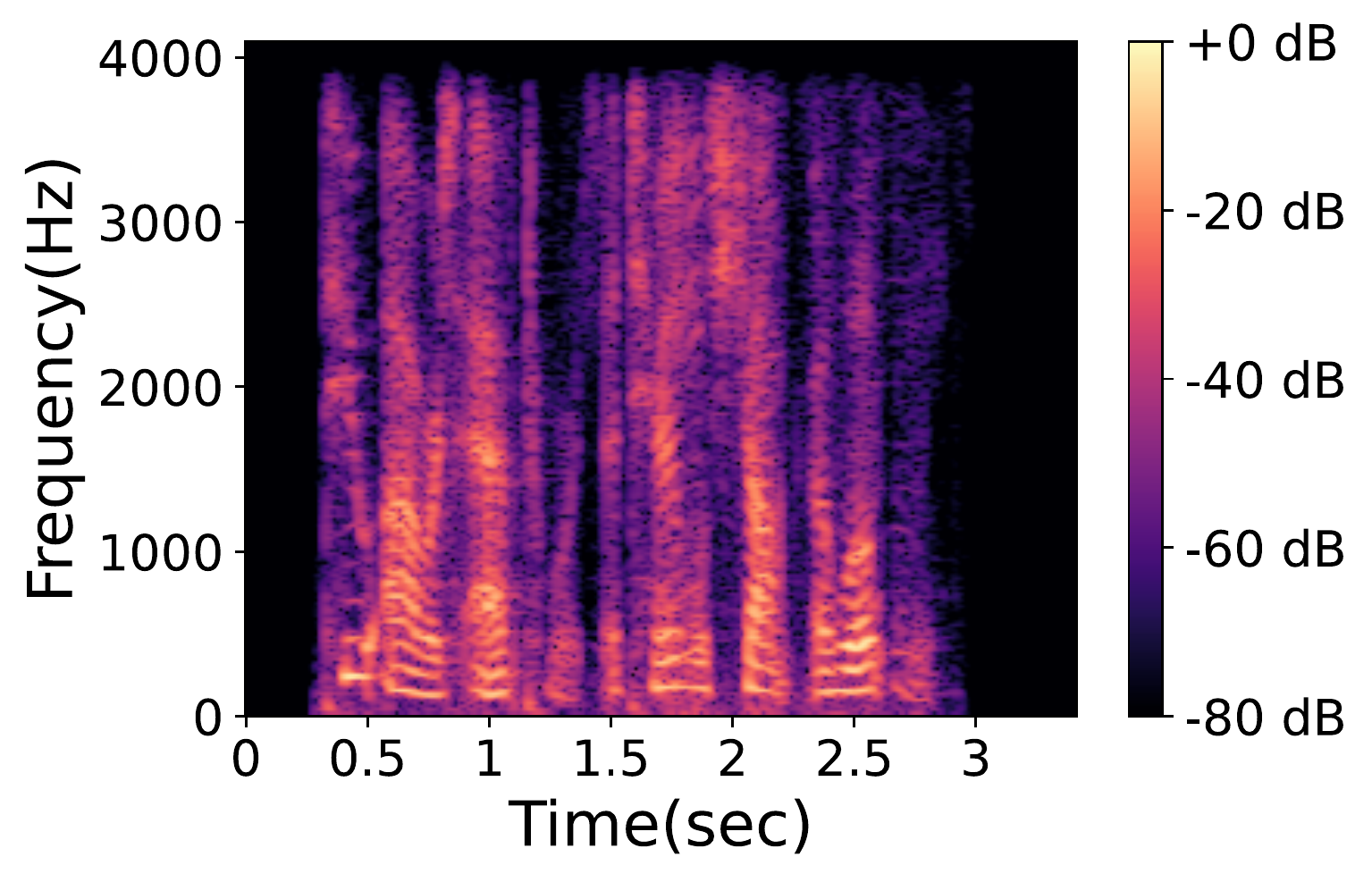}
        \caption{Recovered Spk2}
    \end{subfigure}%
    \caption{An example of the effect of fusion technique on separated signals using EEND-SS. (a) Input mixture of two speakers ( spk1 and spk2) with 60\% overlap. (b)(c) Ground truth for separated signals. (d)(e) Estimated separated signals using separation branch output (before fusion). (f)(g) Estimated separated signals by combining with estimated speaker activity from the diarization branch (after fusion).}
    \label{fig:postprocess}
\end{figure}

An example of the effect of fusion on the separated signals is shown in Fig.~\ref{fig:postprocess}. We show ground truth spectrograms for both speakers and input mixture. We also show spectrograms of separated signals before and after multiplying speech activity. We can see by using estimated speech activity, we can improve the separated signal of the first speaker when the diarization module predicts that the speaker is not active. 

Additionally, we tested our proposed method on Libri2Mix \textit{max} mode\footnote{We used the models trained on \textit{min} mode.} to compare the diarization performance with EEND-based models reported in \cite{yang21c_interspeech}.
As shown in Table~\ref{tab:superb}, EEND-SS achieves higher DER compared to model using LMF as input, as well as 10 other models using self-supervised pretraining for feature extraction. 
However, we were not able to reach their performance using HuBERT~\cite{hsu2021hubert} and wav2vec 2.0~\cite{baevski2020wav2vec}, which are reported to achieve high performances for many other speech processing tasks as well~\cite{yang21c_interspeech}. 
This result indicates room for further improvement using self-supervised features instead of LMF, which is left for future work.

\subsubsection{Flexible Number of Speakers}
\begin{table}[t]
  \caption{Experimental results on Libri2Mix \& Libri3Mix mixture dataset. "SCA" stands for Speaker Counting Accuracy.
  }
  \label{tab:libri23mix}
  \centering
\resizebox{0.49\textwidth}{!}{
  \begin{tabular}{ l| c c c c c }
    \toprule
    \multicolumn{1}{c|}{\textbf{Method}} & 
    \multicolumn{1}{c}{\textbf{STOI}}\!\!($\uparrow$) & 
    \multicolumn{1}{c}{$\mathbf{\textbf{SI-SDR}_{i}}$}\!\!($\uparrow$) &
    \multicolumn{1}{c}{$\mathbf{\textbf{SDR}_{i}}$}\!\!($\uparrow$) &
    \multicolumn{1}{c}{\textbf{SCA}}\!\!($\uparrow$) &
    \multicolumn{1}{c}{\textbf{DER}}\!\!($\downarrow$) \\
    \midrule
    Conv-TasNet & 0.756 & 7.66 & 8.71 & -- & --     \\
    EEND-EDA & -- & -- & -- & 86.2 &    10.16   \\
    EEND-SS ($\lambda_1\!=\!0$) & -- & -- & -- & 90.4 & 8.79 \\
    \midrule
    EEND-SS & 0.760 & 9.31 & 7.50 & \multirow{2}{*}{97.9} & \multirow{2}{*}{6.27} \\
    \hspace{1em} + Fusion  & 0.760 & \textbf{9.38} & 7.59 & &  \\
    \hspace{1em} + LMF & \textbf{0.767} & 8.83 & 9.72 & \multirow{2}{*}{\textbf{98.2}} & \multirow{2}{*}{\textbf{6.04}}  \\
    \hspace{1em} + LMF + Fusion & \textbf{0.767} & 8.87 & \textbf{9.77} & &  \\
    \bottomrule
  \end{tabular}
  }
\end{table}

Next, we evaluate our method on the 2 \& 3-speaker mixed condition created by combining both Libri2 \& 3 Mix datasets.
We followed the training procedure of a flexible number of speakers in \cite{horiguchi2020Interspeech}, and finetuned the models from the weights trained on Libri2Mix. 
In this experiment, the number of reference speech signals $C$ and the separated speech signals $\hat{C}$ may differ due to speaker counting error. To evaluate the separation performance in such cases, we append $|C-\hat{C}|$ silent audio signals to the reference or the separated speech signals to match the number of signals. To avoid the objective metrics from diverging, signals with an amplitude of $10^{-6}$ are used in our implementation.
Since Conv-TasNet cannot perform speaker counting, we use the oracle numbers during inference. 
SCA was also measured in this experiment.

The results for flexible numbers of speakers are shown in Table~\ref{tab:libri23mix}. Likewise, the results for fixed numbers of speakers, EEND-SS outperformed the baseline methods in all the metrics, including speaker counting. Interestingly, EEND-SS also outperforms Conv-TasNet, which uses using oracle number of speakers where EEND-SS estimates the number of speakers. We can assume that EEND-SS learns TCN bottleneck features that are suitable for speech separation as well as speaker diarization thanks to the joint training framework. Thus, we show the effectiveness of joint speech separation, speaker diarization, and speaker counting based on the proposed method for flexible numbers of speakers.

\begin{figure}[t]
    \centering
    \begin{subfigure}[b]{0.244\textwidth}
        \centering
        \includegraphics[width=1.\linewidth]{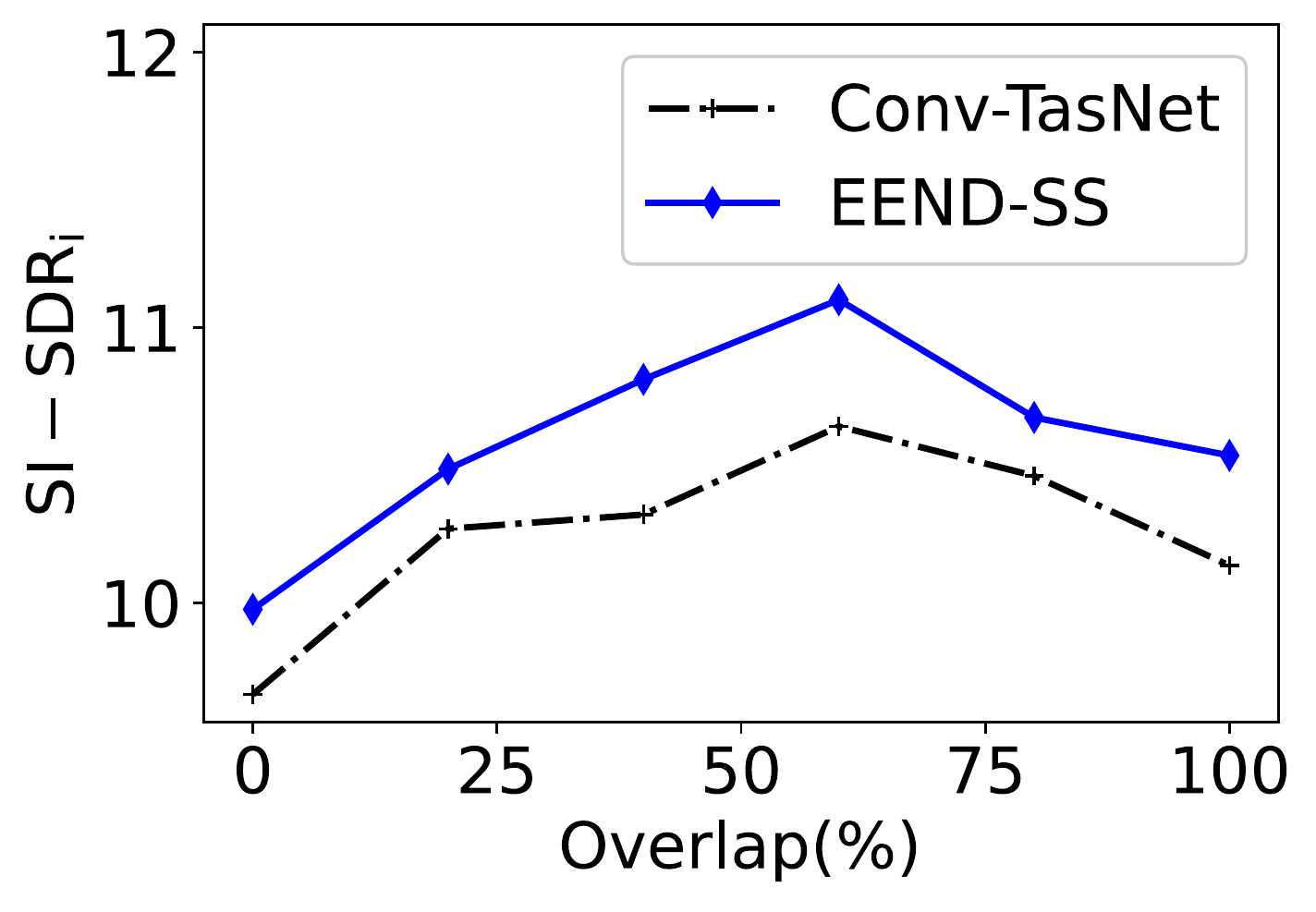}
        \caption{$\text{SI-SDR}_{i}$ ($\uparrow$)}
    \end{subfigure}%
    ~
    \begin{subfigure}[b]{0.244\textwidth}
        \centering
        \includegraphics[width=1.\linewidth]{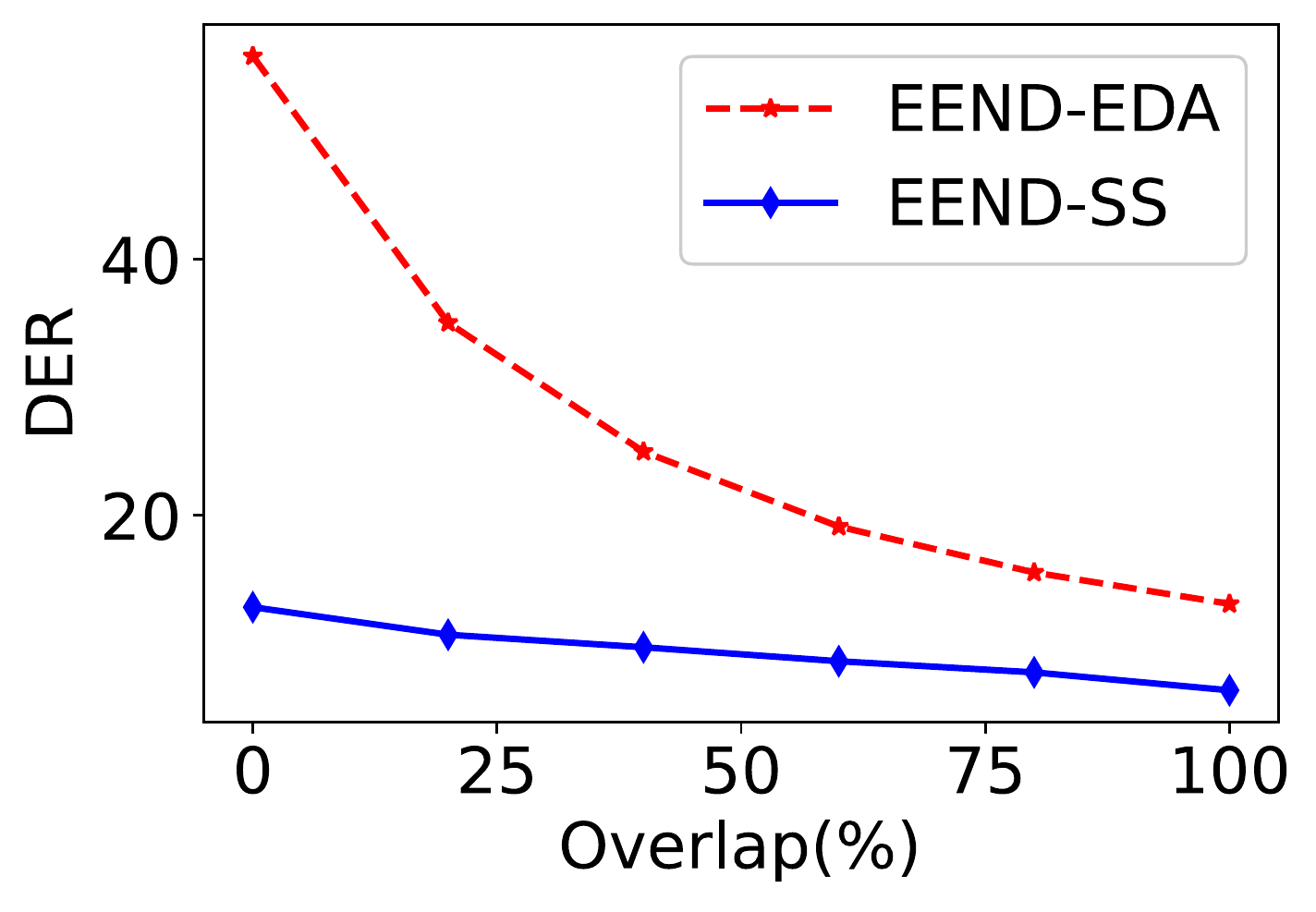}
        \caption{DER ($\downarrow$)}
    \end{subfigure}
    \caption{Experimental results on noisy SparseLibri2Mix. All models are trained with Libri2Mix min mode.}
    \label{fig:sparse_noisy}
\end{figure}

\begin{figure}[t]
    \centering
    \begin{subfigure}[b]{0.244\textwidth}
        \centering
        \includegraphics[width=1.\textwidth]{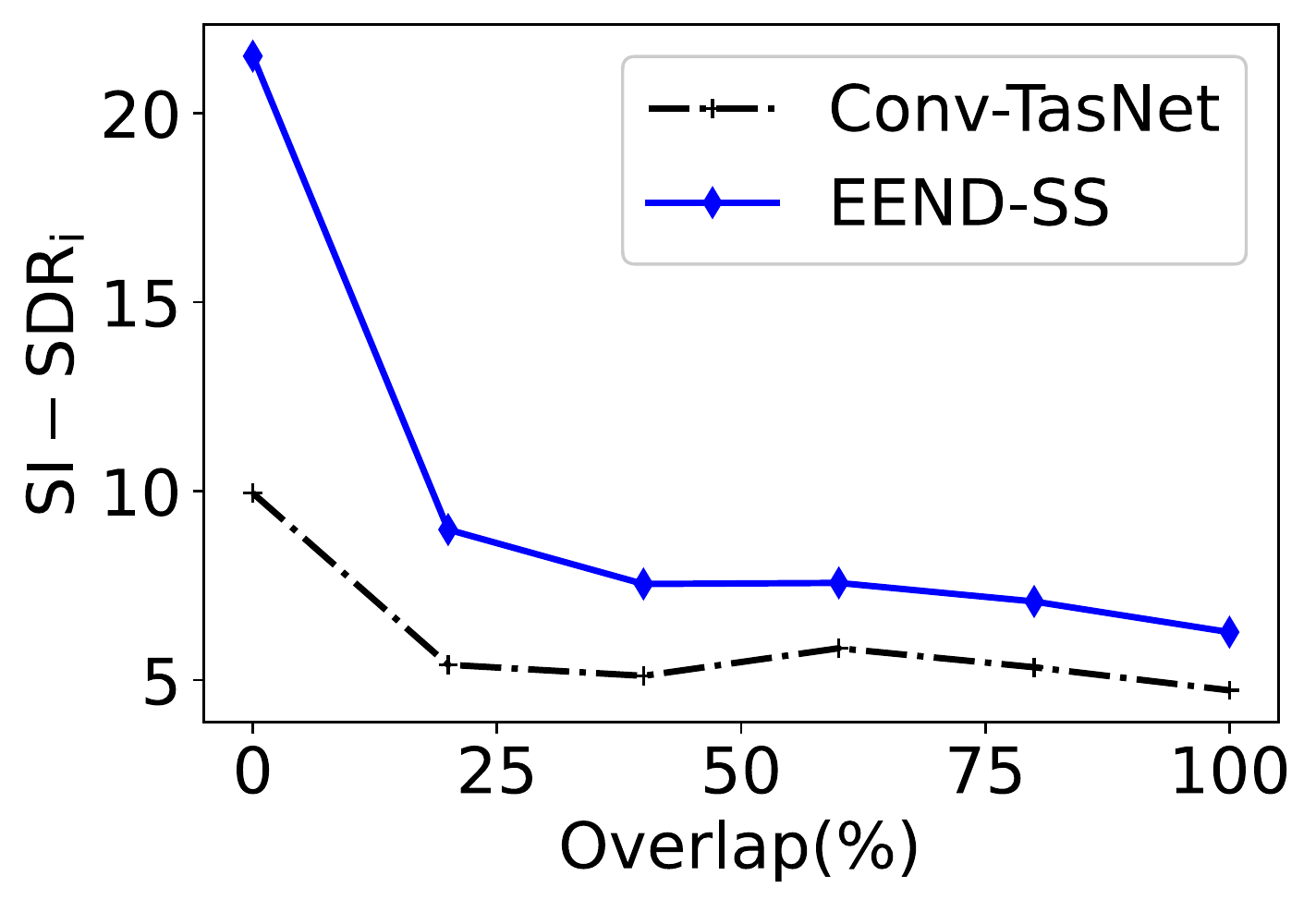}
        \caption{$\text{SI-SDR}_{i}$ ($\uparrow$)}
    \end{subfigure}%
    ~
    \begin{subfigure}[b]{0.244\textwidth}
        \centering
        \includegraphics[width=1.\textwidth]{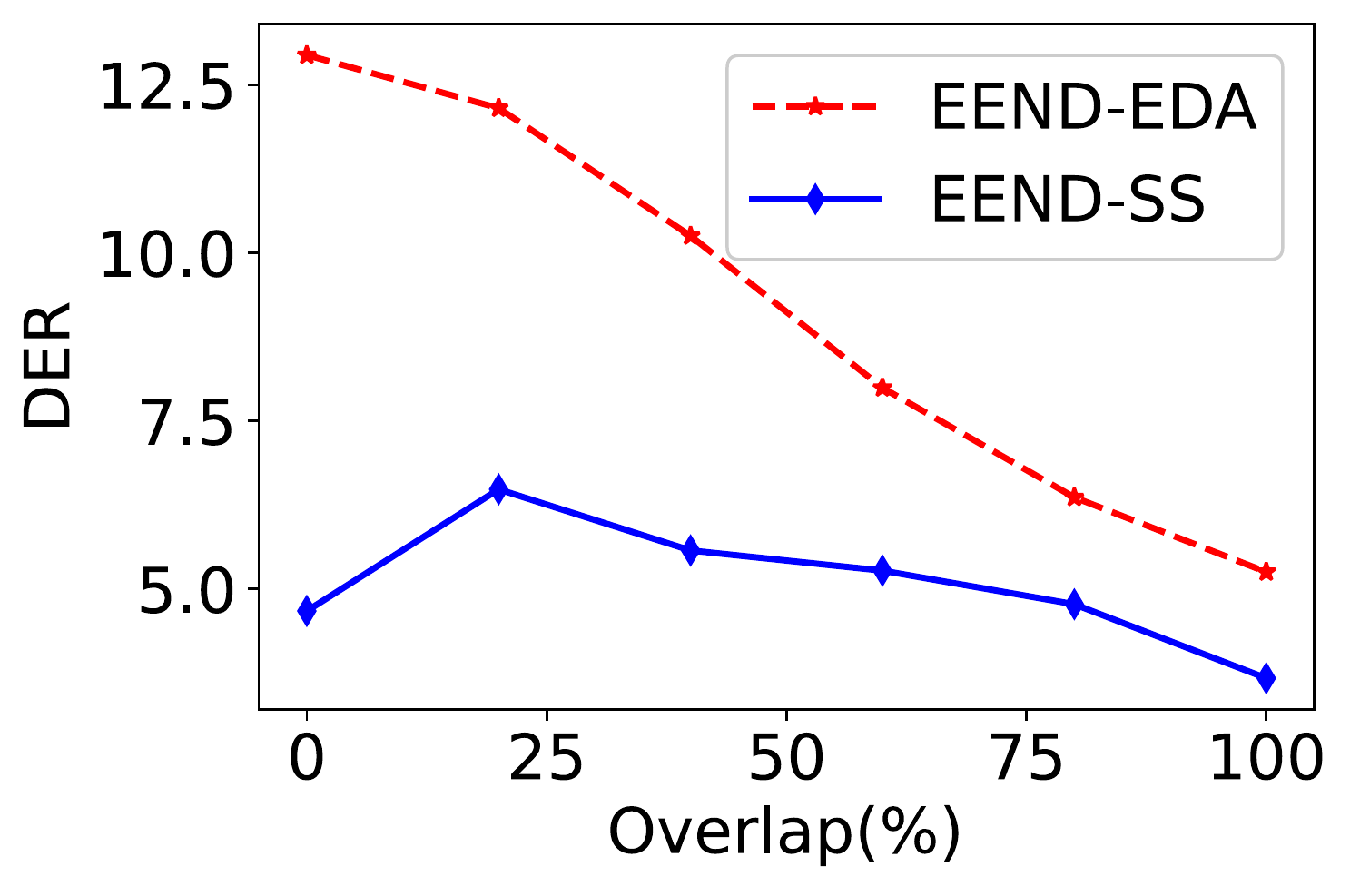}
        \caption{DER ($\downarrow$)}
    \end{subfigure}
    \caption{Results on clean SparseLibri2Mix. All models are trained with simulated SparseLibri2Mix training mixtures.}
    \label{fig:sparse_clean}
\end{figure}

\subsubsection{Sparse Speech Mixtures}
Lastly, we evaluate EEND-SS on sparsely overlapped mixtures. In conversation-like real-world speech mixtures, overlap ratios are typically smaller, speech mixtures are not fully overlapped. For example in meetings~\cite{carletta2005ami} speaker overlap is in the order of 20\% . Sparsely overlapped datasets are better suited for diarization task as well. Since for a fully overlapped mixture predicting speaker activity can be easier, when speaker activity is sparse diarization task is more interesting. We evaluate two cases. First, we evaluate generalizability on sparse-overlapped mixtures with models trained on fully overlapped data, and second, we evaluate the performance of the EEND-SS when trained with a sparse dataset. We compare EEND-SS with EEND-EDA and Conv-TasNet. Here, we use a simulated sparse mixture testset from SparseLibr2iMix~\cite{snyder2015librimix} for our evaluation. SparseLibri2Mix testset contains $2$-speaker mixtures with varying overlap ratio as 0\%, 20\%, 40\%, 60\%, 80\%, and 100\% with $500$ instances for each overlap ratio.

We use fully overlapped Libri2mix trained models (as previously mentioned in Table~\ref{tab:libri2mix}) and test on noisy SparseLibri2Mix mixtures to test the generalization ability of EEND-SS on different speaker overlap than seen in training. We compare the best performing EEND-SS+LMF+Fusion model and compare it with two baselines, ConvTasNet and EEND-EDA. Experimental results for different overlaps are shown in Fig.~\ref{fig:sparse_noisy}. EEND-SS consistently performs better than separation baseline (ConvTasNet) in $\text{SI-SDR}_{i}$ for all overlap scenarios. More interestingly, EEND-SS significantly outperforms EEND-EDA in DER, especially in less overlapped mixtures. We can assume that learning to both separate and diarize is helping the joint model to perform better in diarization when the overlapping scenario is mismatched. Thus, we can say such a multitasking framework can help in better generalization. 


Next, we evaluate on sparse overlapped test sets when EEND-SS has seen such overlap in training. We choose SparseLibri2Mix-clean version, as sparse-overlapped training data generation scripts are available for clean mixtures only. 
We generate train-set with $5000$ mixtures for each of six overlap ratios: 0\%, 20\%, 40\%, 60\%, 80\%, and 100\%, i.e. $30,000$ mixtures in total. We train Conv-TasNet, EEND-EDA, and EEND-SS on the sparse training dataset and report on Fig.~\ref{fig:sparse_clean}.
EEND-SS outperforms baselines in both separation and diarization tasks. EEND-SS performs significantly better than both single-task models in smaller overlapped conditions. Thus, we can say that jointly integrating separation and diarization improves both tasks significantly for sparse-overlapped mixtures. Since sparse mixtures are more suited for diarization task, we can assume that diarization branch can learn important information from the sparse mixture.

\section{Conclusion}\label{sec:conclusion}
In this paper, we proposed a framework to integrate speaker counting, speaker diarization, and speech separation. To enhance the speech separation model, we propose the multiple 1$\times$1 convolutional layers for estimating separation masks for a variable number of speakers and a fusion technique for refining separated speech with estimated speech activity from the diarization branch. We show using LibriMix that the joint framework outperforms single tasks in both fixed and flexible numbers of speakers. Furthermore, we show joint framework improves performance in different overlap scenarios. Future work includes using other separation techniques, as well as using the features from self-supervised pretrained models.

\section{Acknowledgements}
We thank Shota Horiguchi (Hitachi, Ltd.) and Samuele Cornell for their helpful advice. This work used the Extreme Science and Engineering Discovery Environment (XSEDE)~\cite{xsede}, which is supported by NSF grant number ACI-1548562. Specifically, it used the Bridges system~\cite{nystrom2015bridges}, which is supported by NSF award number ACI-1445606, at the Pittsburgh Supercomputing Center (PSC).

\bibliographystyle{IEEEbib}
\bibliography{strings,refs}

\end{document}